\documentclass[a4paper,aps,prd,twocolumn,groupedaddress]{revtex4-2}
\usepackage{natbib}
\usepackage{dcolumn}% Align table columns on decimal point
\usepackage{mathtools}
\usepackage{amsmath}
\usepackage{amssymb}
\usepackage{amsfonts}
\usepackage{mathrsfs}
\usepackage[caption=false]{subfig}
\usepackage{graphicx}
\usepackage{bm}
\usepackage{xcolor} %coloured fonts
\usepackage{bigstrut}
\usepackage{tabularx}
\usepackage{upgreek}
\usepackage{subfig} %multiple Figures per Figure environment
\usepackage{epstopdf} %converts .eps or .ps files to pdf format for Figures
\usepackage[colorlinks=true,citecolor=blue,linktoc=all]{hyperref}
  {\color{red}}%
  {}

\usepackage[normalem]{ulem}  % \sout{old text} for strikeout  

\begin{document}

\title{Two first-order phase transitions in hybrid compact stars: Higher-order multiplet stars, reaction modes and intermediate conversion speeds}

% The list of authors, and the short list which is used in the headers.
% If you need two or more lines of authors, add an extra line using \newauthor
\author{Peter B. Rau}
 \affiliation{Institute for Nuclear Theory, University of Washington, Seattle, Washington, 98195-4550 U.S.A.}
 \email{prau@uw.edu}%Lines break automatically or can be forced with \\
\author{Armen Sedrakian}%
\affiliation{Frankfurt Institute for Advanced Studies, 60438 Frankfurt am Main, Germany\\
Institute of Theoretical Physics, University of Wroc\l{}aw,  50-204 Wroc\l{}aw, Poland}

% These dates will be filled out by the publisher
\date{\today}

\begin{abstract}
We study compact stars with hybrid equations of state consisting of a nuclear outer region and two nested quark phases, each separated from the lower density phase by a strong first-order phase transition. The stability of these models is determined by calculating their radial oscillation modes with different conversion rates between adjacent phases and hence junction conditions for the modes at the phase separation interface between them. In the case when the timescale of transition is faster than the period of oscillations, we recover the traditional stability criterion implying 
that $\partial M/\partial\rho_c>0$ on the stable branch(es), where $M$ is the mass and $\rho_c$ is the central density. In the opposite limit of slow conversion, we find  stable stellar multiplets beyond triplets consisting of stars that are stable by the usual  criterion plus  slow-conversion (denoted by $s$) hybrid stars  with $\partial M/\partial\rho_c<0$ that are stabilized due to an alternative junction condition on the  fluid displacement field at the interface reflecting the slow rate of conversion.  We also study the properties of the reaction mode, the radial mode that only exists for stars with rapid (abbreviated by $r$) phase transitions, in stars with either two rapid phase transitions or alternating rapid and slow phase transitions for the first time. The implications of alternative junction conditions are also examined, with these conditions generally being found to provide stability properties similar to those for a slow conversion rate.
\end{abstract}

\maketitle

\section{Introduction}

The study of various classes of compact stars that are distinguished by the composition of their dense cores is one of the best ways to constrain the equation of state (EoS) of strongly interacting matter at densities above nuclear saturation. Compact star models with quark cores -- {\it hybrid stars} -- are of great interest because they probe the properties of the phase diagram of  dense quantum chromodynamics (QCD) via astrophysical observations. (For recent reviews see~\cite{Alford2019JPhG,Baym2018RPPh}).

%These are often contrasted with the hypothesis of strange stars - a distinct class of stars based on the hypothesis that the strange matter is the true ground state of  matter in the Universe~\cite{}.

Oscillations of stars are known to be a sensitive probe of their internal composition.
For example, the recent studies of the spectrum of modes  in the case of compact stars show that higher $g$-mode frequencies are expected in their cores containing quark matter  as opposed to purely hadronic matter~\cite{Wei2020,Jaikumar2021}. The $g$-modes are also sensitive to the presence of and size of the density jump at a first-order phase transition between hadronic and quark matter~\cite{Orsaria2019}. Hence the resulting phase shift in the emitted gravitational wave signal due to the excitation of $g$-modes during the binary inspiral of a neutron star with another neutron star or black hole could indicate the presence and properties of this transition. Even if the density of this phase transition is not attained in stable compact stars, signatures of a phase transition to quark matter could appear during the mergers of binaries composed of two compact stars, since densities even higher than those present in stable stars can be reached. A phase transition to deconfined quark matter shortly after the merger may leave a signature in the gravitational wave and electromagnetic signals from the postmerger object~\cite{Bauswein2019,Weih2020,Liebling2021,Prakash2021}, although it is not clear whether these shifts will not be overshadowed by other effects.

Despite substantial progress in our understanding of QCD both through lattice studies and phenomenological models, the nature of the deconfinement transition from nuclear matter to phase(s) of quark matter remains an open question. The deconfinement can occur via a first-order phase transition with a density discontinuity~\cite{Fukushima2011,Schmidt2017}, a smooth hadron-quark crossover~\cite{Masuda2013a,Baym2019ApJ,Minamikawa2021}, or a second-order transition from nuclear matter to quarkyonic matter~\cite{McLerran2019}. 
The deconfined quark matter phase diagram may contain various phases depending on strange quark mass, beta-equilibrium conditions, chiral symmetry,  and its breaking, etc.  Examples of phases include (but are not limited to) various superconducting phases of quarks~\cite{Alford2019JPhG} or quarkyonic matter~\cite{McLerran2019}. Although the various phases differ primarily by their spectrum close to the Fermi surface(s) of quarks, their specific features - for example, the gapped spectrum - affect the global structure of compact stars~\cite{Alford2003,Alford2005}. 

An observationally testable conjecture that follows from first-order phase transition is  the existence of twin stars --
hadronic stars and hybrid stars with the same gravitational mass but different radii and internal compositions~\cite{Gerlach1968,Kampfer1981,Glendenning2000,Schertler2000,Alford2013,Benic2015,Li:2019fqe,Li:2021sxb,Alvarez-Castillo2019,Christian2021,Dexheimer2021JPhG}.  Note that a first-order phase  transition is required for twin stars to exist, but it does not guarantee their existence, as  it could also lead to a collapse to a black hole. In the case of two strong first-order phase transitions (one of which is within the quark matter itself), triplets of stars can arise - three stars with the same masses but different radii~\cite{Alford2017a}. An example of such transition is that from 2SC to CFL color superconducting phase~\cite{Alford2008,Bonanno:2011ch}. The most compact star in a triplet contains two phases of quark matter, the next most compact star - has a single phase, and the least compact star is purely hadronic. 

The dynamical stability of compact stars is also an important question, as an unstable star will quickly collapse into a black hole. One of the most common methods to determine the stability of a star is by requiring that its mass must increase with increasing central density (or central pressure) $\partial M/\partial\rho_c>0$. This is part of the Bardeen--Thorne--Meltzer (BTM)~\cite{Bardeen1966} stability criterion. Although violations of this criterion have been known to exist numerically for decades~\cite{Gourgoulhon1995} (also in the case of rotating stars, see Ref.~\cite{Takami2011MNRAS}), it  is still often used as it only requires solving the Tolman--Oppenheimer--Volkoff (TOV) equation for a given EoS. A more fundamental criterion for stability is that the radial modes of oscillation of a star have real frequencies. Since calculating the oscillation modes is a Sturm--Liouville problem, they can be arranged such that their successive eigenvalues (i.e., frequencies squared) must increase. Hence, if the lowest-order radial mode, with zero nodes (i.e., the fundamental or $f$-mode) has a positive frequency-squared, the stellar configuration is dynamically stable; otherwise, it is unstable. This criterion requires solving for the oscillation modes of the star in addition to computing the background stellar models using the TOV equation. It is generally found that, for a single phase of dense matter, both stability criteria agree.  

The study of stellar oscillations of compact stars with phase transitions was pioneered by~\cite{Bisnovatyi-Kogan1984}, with a nonrelativistic star composed of two phases of uniform density, and~\cite{Haensel1989}, who studied nonrelativistic stellar models with a phase transition and non-uniform density profiles. These works emphasized the importance of the rate of the phase transition: whether or not a fluid element at the phase transition preserves its phase as the oscillation carries it across the equilibrium boundary between the two phases, which is related to the rate of the phase transition compared to the oscillation period. If fluid elements perturbed across this boundary immediately undergo a phase transition (a \textit{rapid} transition), there emerges a distinct class of oscillation mode, the \textit{reaction mode}~\cite{Haensel1989}. The rate of a phase transition may additionally have physically-relevant implications for the stability of compact stars as shown in Ref.~\cite{Pereira2018}. Specifically, they showed that if the conversion rate is slow, i.e., its characteristic timescale is longer than the period of oscillations then the stability of the fundamental mode extends beyond the maximum of the $M(\rho_c)$ curve, i.e., stars on the $\partial M/\partial\rho_c<0$ branch are stable against $f$-mode oscillations. Such stars  were coined as {\it slow stable} or $s$ hybrid stars~\cite{Curin2021,Lugones2023}. Recently, 
Ref.~\cite{Goncalves2022} extended the stability analysis to EoS with two-phase transitions and examined how the stability conditions change when there are two distinct phases in quark matter. 

In this paper, we extend the study of the stability of compact stars with two-phase transitions in several directions. We first employ EoSs that support classical twin and triplet star configurations. By assuming slow conversion we then look for higher-order multiplet stars
due to the fact that $s$ hybrid stars increase the number of stable stars with the same masses for any given EoS (for example from a single star to a twin, from twin to triplet, from triplet to a quadruplet, etc.) We then examine the reaction mode of a compact star with one or more rapid phase transitions, which were not studied in Ref.~\cite{Goncalves2022}. We conclude by examining different choices of junction conditions for the fluid perturbations across the phase transitions besides the usual slow ($s$) or rapid ($r$) cases and examine their implications for dynamic stability.

In Sections~\ref{sec:EoS} and~\ref{sec:ModeFreqs} we review the equations of state that we use in the paper and the calculation of the radial modes of a relativistic star. In Section~\ref{sec:Multiplets} we demonstrate the existence of potential higher-order multiplet stars composed of classical twin and triplet configurations plus $s$-stable stars. In Section~\ref{sec:ReactionMode} we discuss the reaction modes for a star with multiple first-order phase transitions. Section~\ref{sec:Junctions} discusses alternative junction conditions to the usual fast and slow conditions, including more physically realistic conditions. The results are reviewed in Section~\ref{sec:Conclusion}. We work in units where $c=G=1$ throughout the paper.

\section{Equations of state}
\label{sec:EoS}

Current models of hybrid stars are commonly constrained by the multimessenger data which includes massive pulsars~\cite{Demorest2010,Antoniadis2013,Cromartie2020}, the GW170817 gravitational wave event~\cite{LIGO_Virgo2017a} and NICER X-ray observations~\cite{Riley2019,Miller2019,Riley2021,Miller2021}. Below, the model parameters that will be used are consistent with these constraints within $2\sigma$ confidence intervals.

\begin{figure}
\includegraphics[width=\columnwidth]{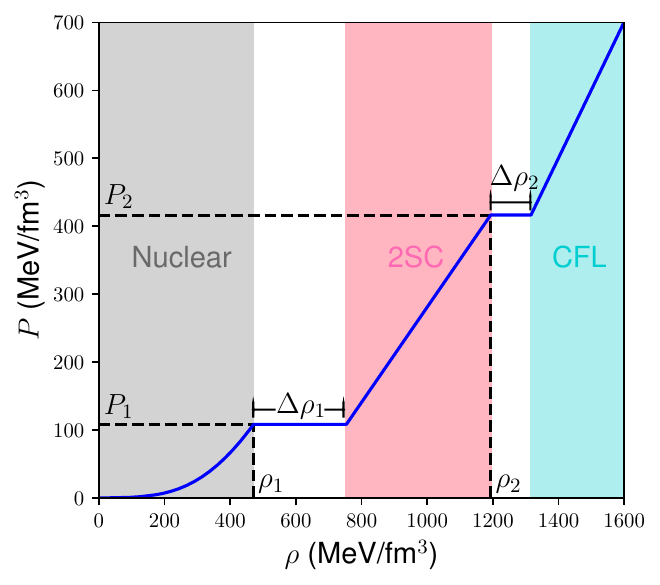}
\caption{The EoS for parametrization A showing the definitions of the parameters in Eq.~(\ref{eq:EOS}) and labelling the three distinct phases.}
\label{Fig:EOSParamA}
\end{figure}

Below, we use the same zero temperature, chemically equilibrated, constant speed of sound equations of state as in Ref.~\cite{Alford2017a} that describe compact stars with two first-order phase transitions. The three phases of matter are taken to be nuclear matter and two distinct quark matter phases, for instance, two different color-superconducting phases like 2SC (two-flavor color-superconducting) and CFL (color-flavor-locked) phases. We use this nomenclature for the two quark-matter phases, though we do not claim that these phases must be color-superconducting. The nuclear matter EoS is the DDME2 EoS from~\cite{Colucci2013},  which is calculated using relativistic density functional theory. The different EoSs are then parametrized by choosing six parameters: 1) the energy density $\rho_1$ at the first (nuclear to 2SC) phase transition, which along with the DDME2 EoS gives the pressure $P_1$ at the lower density edge of the nuclear--2SC phase transition; 2) the energy density $\rho_2$ at the lower-density edge of the 2SC--CFL transition; 3) and 4) the energy density jumps of the two phase transitions $\Delta\rho_1$ and $\Delta\rho_2$; 5) and 6) the two constant sound speeds squared $c_{s1}^2$ and $c_{s2}^2$ in the 2SC and CFL phases. The pressure $P_2$ of the 2SC to CFL phase transition is determined by $P_2=P_1+c_{s1}(\rho_2-\rho_1-\Delta\rho_1)$. Hence the EoS is expressed as
\begin{equation}
P(\rho)=\begin{cases} 
       P_{\text{DDME2}}(\rho), & \rho<\rho_1, \\
      P_1, & \rho_1\leq\rho\leq\rho_1+\Delta\rho_1, \\
      P_1 + c_{s1}^2\left[\rho-\rho_1-\Delta\rho_1\right], & \rho_1+\Delta\rho_1<\rho<\rho_2, \\
      P_2, & \rho_1\leq\rho\leq\rho_2+\Delta\rho_2, \\
      P_2 + c_{s2}^2\left[\rho-\rho_2-\Delta\rho_2\right], & \rho_2+\Delta\rho_2<\rho.
   \end{cases}
   \label{eq:EOS}
\end{equation}
We choose $c_{s1}^2= 0.7c^2$ or $0.8c^2$, and $c_{s2}^2=c^2$ for all calculations, but allow the other four parameters $P_1$, $\rho_2$, $\Delta\rho_1$, $\Delta\rho_2$ to vary. We are using the Maxwell construction to join the different phases (hence the large density jumps), but other choices (the Gibbs construction or a crossover transition) are in principle possible. This EoS, for parametrization A (see Table~\ref{tab:EoSParameters} later in the paper for the specific parameters) is show in Figure~\ref{Fig:EOSParamA}.

Ref.~\cite{Alford2017a} shows that the above choice of the EoS with two phase transition leads to classically (i.e., in the case of rapid conversion) stable twin and triplet compact star configurations for some range of parameters. For our purposes, such parameter sets are  of particular interest for studies of the appearance of higher-order multiplets of compact stars such as quadruplets, quintuplets, and sextuplets. Such multiplets can indeed exist because one needs to add to classically stable stars those stars on the classically unstable branch which are stabilized by the conjecture of slow conversion. 

\section{Calculation of the mode frequencies}
\label{sec:ModeFreqs}

The formalism to compute the radial normal modes of a general relativistic star was first developed by Chandrasekhar in  Ref.~\cite{Chandrasekhar1964a}.  Below, we use an alternative set of equations given by ~\cite{Chanmugam1977,Gondek1997}
\begin{align}
\frac{\textrm{d}\xi}{\textrm{d}r}={}&\left(\frac{\textrm{d}\nu}{\textrm{d}r}-\frac{3}{r}\right)\xi-\frac{\Delta P}{r\Gamma P},
\label{eq:XiEq}
\\
\frac{\textrm{d}\Delta P}{\textrm{d}r}={}&\left[\textrm{e}^{2\lambda}\left(\omega^2 \textrm{e}^{-2\nu}-8\pi P\right)+\frac{\textrm{d}\nu}{\textrm{d}r}\left(\frac{4}{r}+\frac{\textrm{d}\nu}{\textrm{d}r}\right)\right]
\nonumber
\\
{}&\times\left(\rho+P\right)r\xi
\nonumber
\\
{}&-\left[\frac{\textrm{d}\nu}{\textrm{d}r}+4\pi(\rho+P)r\textrm{e}^{2\lambda}\right]\Delta P,
\label{eq:DeltaPEq}
\end{align}
where $\xi=\xi_{\rm dim}/r$ with $\xi_{\rm dim}$ being the standard dimensionful definition of the Lagrangian displacement and $r$ being the radial coordinate, $\Delta P$ is the Lagrangian perturbation of pressure  $P$,  $\rho$ is mass-energy density, $\omega$ is the angular frequency, $\Gamma$ is the adiabatic index, and $\text{e}^{2\nu}$ and $\text{e}^{2\lambda}$ are metric coefficients for the interior metric tensor of a spherically symmetric, static perfect fluid. We note that Eq.~\eqref{eq:XiEq}--\eqref{eq:DeltaPEq} are derived without using the Cowling approximation.

The metric tensor is given by
\begin{equation}
\text{d}s^2 = -\text{e}^{2\nu}\text{d}t^2+\text{e}^{2\lambda}\text{d}r^2+r^2(\text{d}\theta^2 + \sin^2\theta \text{d}\phi^2).
\end{equation}
$\lambda$ and $\nu$ are functions of $r$ only. Like $\rho(r)$ and $P(r)$, they are determined by solving the TOV equations for a particular central pressure $P_c$ and choice of EoS. The TOV equations are given by
\begin{align}
\frac{\text{d}P}{\text{d}r}={}&-\frac{1}{r^2}\frac{(\rho+P)(m+4\pi r^3P)}{(1-2m/r)},
\\
\frac{\text{d}m}{\text{d}r}={}&4\pi\rho r^2,
\end{align}
where $m$ is the enclosed by $r$ gravitational mass. They  
 are integrated from $r=0$ and $P=P_c$ outward until $P=0$, which then fixes the stellar  radius $r=R$ and the gravitational mass $M=m(r=R)$ of the star. The functions $\lambda(r)$ and $\nu(r)$ are given by
\begin{align}
\lambda(r)={}&-\frac{1}{2}\ln\left(1-\frac{2m(r)}{r}\right),
\\
\nu(r)={}&-\int^r_0\frac{\text{d}P}{\rho+P}+C,
\end{align}
where $C$ is a constant determined by requiring that $\nu(r=R)$ matches the Schwarzschild metric at the stellar exterior 
\begin{equation}
\nu(r=R)=\frac{1}{2}\ln\left(1-\frac{2M}{r}\right).
\end{equation}
The adiabatic index for a chemically equilibrated relativistic fluid is
\begin{equation}
\Gamma = \frac{\rho+P}{P}\left(\frac{\partial P}{\partial\rho}\right)_{s},
\end{equation}
where the entropy per particle $s$ is held constant. In the nuclear matter region of the EoS, $\Gamma$ is computed by finite-differencing $P(\rho)$, while in the quark matter phases, it is computed using the constant sound speeds squared $c_s^2=(\partial P/\partial\rho)_{s}$. 
Following~\citep{Pereira2018}, we make the approximation that the matter is always in chemical equilibrium (corresponding to beta equilibrium in the nuclear matter) since the difference between the $\Gamma$ calculated in beta equilibrium and with frozen composition differ by $\lesssim15$\% in the relevant density and temperature range~\citep{Haensel2002a}.

At the center of the star, we require the displacement field to be divergence-free, which according to Eq.~(\ref{eq:XiEq}) implies
\begin{equation}
\Delta P(r=0) = -3\Gamma P\xi(r=0).
\end{equation}
The overall amplitude of $\xi$ is unconstrained and we choose the normalization $\xi(r=0)=1$.
This leaves us with a single free parameter -- the  mode frequency squared $\omega^2$. As a boundary condition we require that $\Delta P(r=R)=0$ for all allowed frequencies $\{\omega_i\}$. We use a shooting method to compute $\omega^2$, i.e.,   starting from $r=0$, Eqs.~\eqref{eq:XiEq} and~\eqref{eq:DeltaPEq} are integrated outward for a given value of $\omega^2$ until the outer boundary condition is satisfied. The values of $\omega^2$ are labeled according to the number of radial nodes in $\xi$; the stability is determined by the nodeless mode $\omega_0^2>0$.

At the phase transitions between the nuclear and 2SC phases and the 2SC and CFL phases, junction conditions that relate the values of $\xi$ and $\Delta P$ on each side of the transition are needed. These junction conditions were first worked out in the nonrelativistic case in Ref.~\cite{Haensel1989}, and were later generalized to the general relativistic case in Ref.~\cite{Karlovini2004}. These conditions then were employed in Refs.~\cite{Pereira2018,Goncalves2022} to obtain fundamental modes of hybrid stars with slow and rapid conversion. In the case of a slow conversion rate, these conditions are given by
\begin{equation}
\left[\Delta P\right]^+_-=0, \qquad \left[\xi\right]^+_-=0,
\label{eq:SlowConversionJunctions}
\end{equation}
with the $+$ and $-$ corresponding to the high and low-density sides of the transition. For a rapid conversion rate, the junction conditions are
\begin{equation}
\left[\Delta P\right]^+_-=0, \qquad \left[\xi-\frac{\Delta P}{r}\left(\frac{\textrm{d}P}{\textrm{d}r}\right)^{-1}\right]^+_-=0.
\label{eq:RapidConversionJunctions}
\end{equation}
As the timescale of conversion between hadronic and quark phases (the quark matter nucleation timescale) is highly model-dependent~\citep{Bombaci2016,Lugones2016}, we allow for both rapid and slow phase transitions at the nuclear-2SC phase transition. We do likewise for the 2SC-CFL phase transition in the absence of detailed knowledge of the timescale of such a phase transition. We later modify the above junction conditions to examine the stability implications of alternative junction conditions in Section~\ref{sec:Junctions}.

\section{Higher-order stellar multiplets}
\label{sec:Multiplets}

The potential existence of twin stars consisting of one classically stable neutron star and a slow stable hybrid star was discussed by Ref.~\cite{Pereira2018}. They also  mentioned that higher-order stellar multiplets could exist for EoSs with a third or fourth family of stable stars. Subsequently, Ref.~\cite{Goncalves2022} employed EoSs with two strong first-order phase transitions, which~\cite{Alford2017a} showed can support classically stable triplet stars. However,  they did not use parametrizations of the EoS which support even classically stable twin stars. In this section we examine parametrizations that permit the existence of both classically stable twins and triplets and determine whether they can support additional slow stable hybrid stars, becoming slow stable quadruplets, quintuplets, and sextuplets.

\begin{figure}[tb]
\begin{center}
\includegraphics[width=\columnwidth]{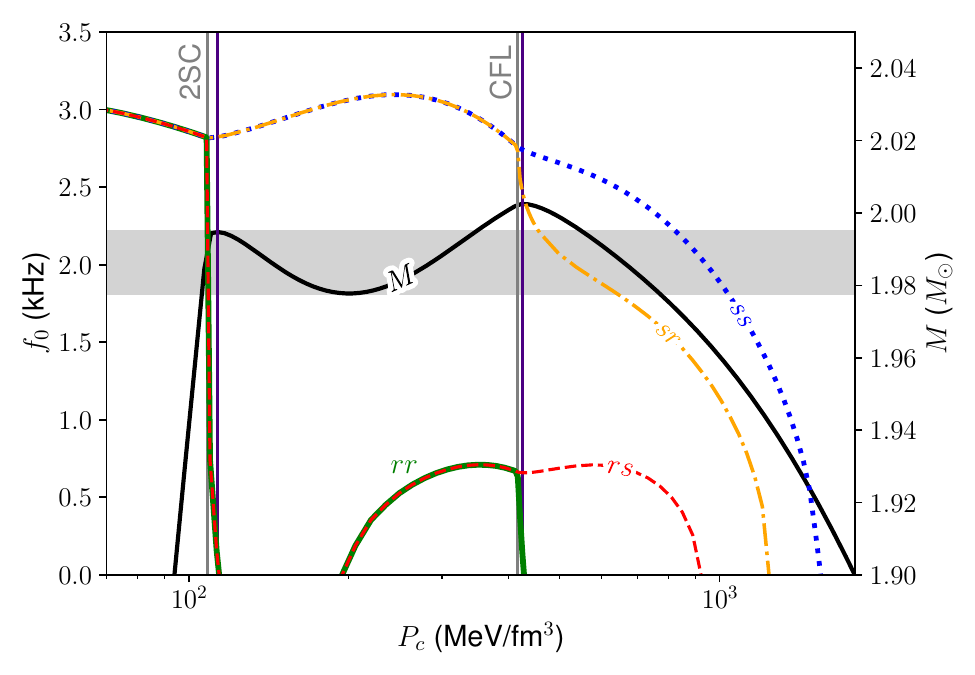}
\end{center}
\caption{Fundamental radial modes $f_0$ as a function of the central pressure $P_c$ of a stellar model for each of the four different configurations of slow ($s$) and rapid ($r$) transitions at the two phase transitions, with lower density phase transition labeled first. The EoS parametrization is given by column A in Table~\ref{tab:EoSParameters}. The pressures of the two phase transitions are the vertical lines labeled 2SC and CFL respectively. The mass $M$ for the given $P_c$ is labeled on the right vertical axis, with the unlabelled vertical lines marking the local maxima of $M(P_c)$. This EoS supports classical twin stars in the shaded mass range. }
\label{Fig:QuadrupletModes}
\end{figure}

Next, following Ref.~\cite{Goncalves2022}, we label the stellar configuration resulting from EoS with the two phase transition by the rates for each phase transition
$ab$ where $a =s$ or $r$ for a slow or rapid nuclear--2SC transition, and where $b=s$ or $r$ for a slow or rapid 2SC--CFL transition. For the EoS parametrization A in Table~\ref{tab:EoSParameters}, we plot the fundamental radial mode (cyclical) frequency $f_0=\omega_0/(2\pi)$ as a function of central pressure $P_c$ in Fig.~\ref{Fig:QuadrupletModes}. The curves are not shown when $f_0$ becomes imaginary/unstable. The mass of the stellar model corresponding to the central pressure $P_c$ is overlaid and its values are shown on the secondary vertical axis. Vertical lines corresponding to local maxima in $M(P_c)$ (violet) and the two-phase transitions (grey) are included. These features are common to each mode spectrum in all Figures in the remainder of this paper.

The EoS parametrization A supports classical twin stars, with the mass range for these twins shaded in the Figure. For two rapid transitions  ($rr$ configuration), only the classical twins are stable as expected, but for the configurations with at least one slow transition one ($rs$) or two ($sr$ and $ss$) additional \textit{slow stable} stellar models are stable in the classical twin mass range, hence the EoS supports a triplet of two classically stable and one slow stable star for the $rs$ configuration, and a quadruplet of two classically stable and two slow stable stars for the $sr$ and $ss$ configurations. Interestingly, the presence of the slow stable nuclear--2SC transition stabilizes the $sr$ configuration beyond the 2SC--CFL transition, which is rapid for this configuration, so that there exist stable stellar models with $P_c$ greater than the 2SC-CFL transition pressure, which is not the case for the $rr$ configuration.

\begin{table}
	\caption{EoS parametrizations used in this paper. Definitions of each parameter are given in the text in Section~\ref{sec:EoS}. Parameters are given in units of MeV/fm$^3$ except for $c_{s1}^2$ and $c_{s2}^2$, which are in units of $c^2$.}
	  \centering
    \begin{tabular}{|c|c|c|c|c|}
    \hline
	\multicolumn{1}{|c|}{} & \textbf{A} & \textbf{B} & \textbf{C} & \textbf{D} \\ 
    \hline
    $\rho_1$ & 471.2 & 384.8 & 471.2 & 471.2 \\
	\hline
	$P_1$ & 108.4 & 59.2 & 108.4 & 108.4 \\
	\hline  
	$\rho_2$ & 1193.2 & 739.8 & 880.7 & 819.9 \\
	\hline  
	$\Delta\rho_1$ & 282.7 & 269.4 & 282.7 & 282.7 \\
	\hline  
	$\Delta\rho_2$ & 122.8 & 230.3 & 116.6 & 115.9 \\
	\hline  
	$c_{s1}^2$ & 0.7 & 0.7 & 0.8 & 0.7 \\
	\hline  
	$c_{s2}^2$ & 1 & 1 & 1 & 1 \\
	\hline   
    \end{tabular}
    \label{tab:EoSParameters}
\end{table}

\begin{figure}
\includegraphics[width=\columnwidth]{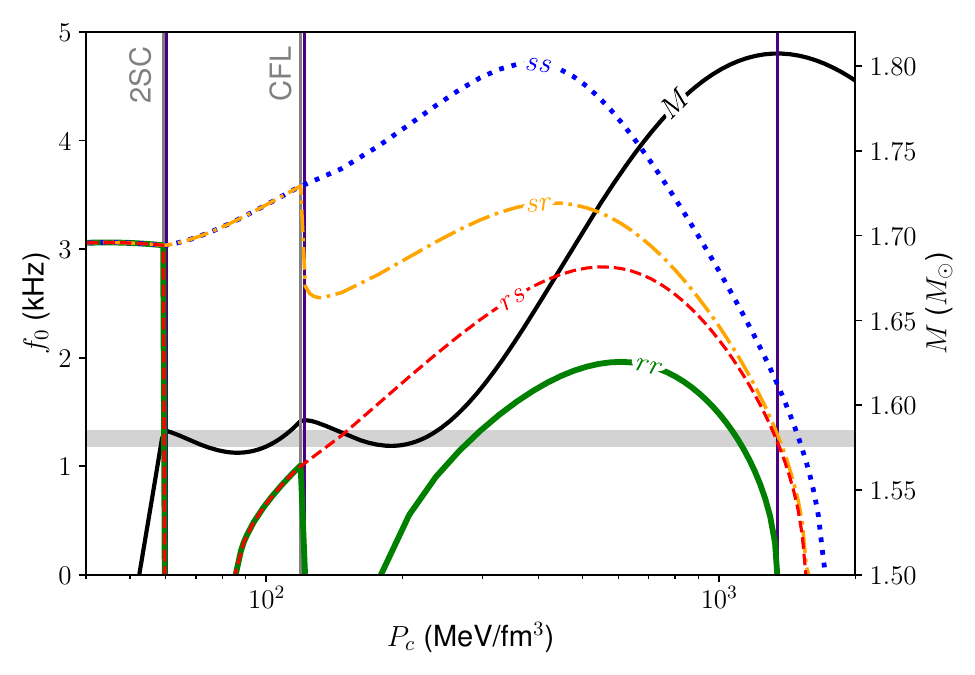}
\caption{Fundamental radial modes $f_0$ for each of the four different configurations of slow $s$ and rapid $r$ transitions at the two phase transitions for EoS parametrization B in Table~\ref{tab:EoSParameters}. This EoS supports classical triplet stars in the shaded mass range, and slow stable quintuplets for a slow nuclear--2SC transition.}
\label{Fig:QuintupletModes}
\end{figure}

In Fig.~\ref{Fig:QuintupletModes}, we plot the fundamental radial modes for the EoS parametrization B in Table~\ref{tab:EoSParameters}, which supports classically stable triplet stars. Like the previous EoS parametrization, for the $rr$ configuration only the classically stable stellar models are stable, whereas one ($rs$) and two ($sr$ and $ss$) additional slow stable stellar models in the classical triplet mass range are supported if there is a slow phase transition inside the star. This EoS, therefore, supports slow stable quintuplets or quadruplets. The maximum mass supported by this parametrization is too low to agree with the astronomical constraint of $2M_{\odot}$, but it does illustrate an important point: the mass does not reach a local maximum and begins to decrease back into the classical triplet mass range until such high central pressures that the fundamental radial modes for the $ss$, $rs$, and $sr$ configurations are imaginary. Thus there is no third slow stable configuration for this EoS.

\begin{figure}
\includegraphics[width=\columnwidth]{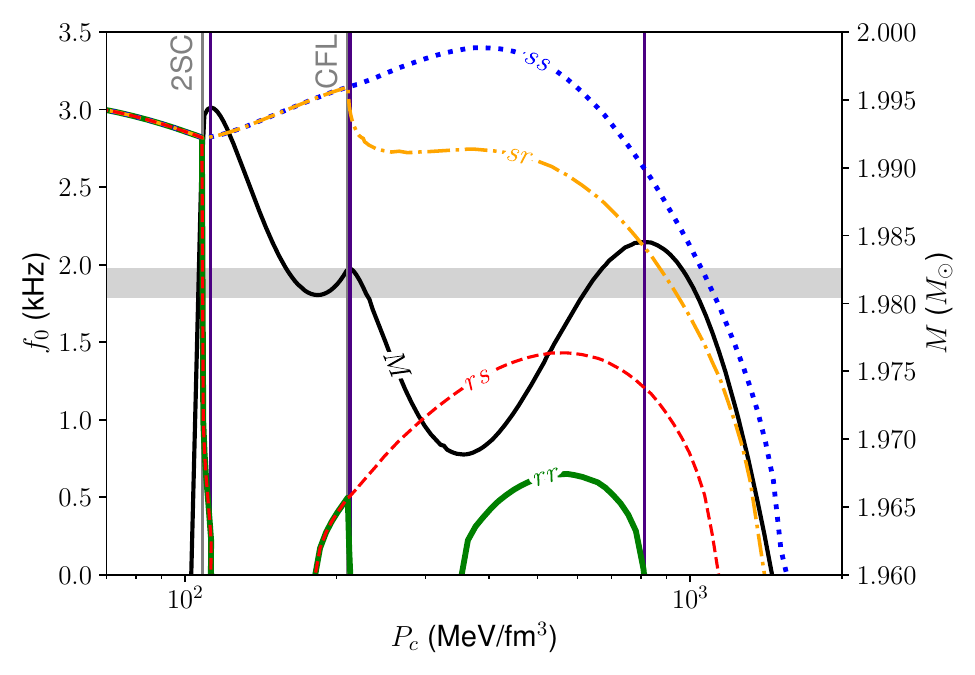}
\caption{Fundamental radial modes $f_0$ for each of the four different configurations of slow $s$ and rapid $r$ transitions at the two phase transitions for EoS parametrization C in Table~\ref{tab:EoSParameters}. This EoS supports classical triplet stars in the shaded mass range, and slow stable sextuplets for a slow nuclear--2SC transition.}
\label{Fig:SextupletModes}
\end{figure}

However, it is possible to find EoS parametrizations that can support sextuplets consisting of three classically stable and three slow stable stars. This requires choosing a parametrization such that the stellar mass has a local maximum for $P_c$ sufficiently close to the 2SC--CFL phase transition that the fundamental radial mode for the slow 2SC--CFL transition case is stable for $P_c$ above the point at which the stellar mass drops below the triplet band above the final local maximum. Fig.~\ref{Fig:SextupletModes} shows the fundamental radial modes for such a parametrization (C in Table~\ref{tab:EoSParameters}). The $rr$ configuration matches the classical result and only supports stable triplet stars. But now $f_0$ for the $ss$, $sr$ and $rs$ configurations becomes imaginary for $P_c$ above that for which the stellar mass passes through the classical triplet mass band for the sixth time, and so slow stable stars are supported here. The $ss$ and $sr$ configurations both support slow stable sextuplets, while the $rs$ configuration supports slow stable quintuplets.

\begin{figure}
\includegraphics[width=\columnwidth]{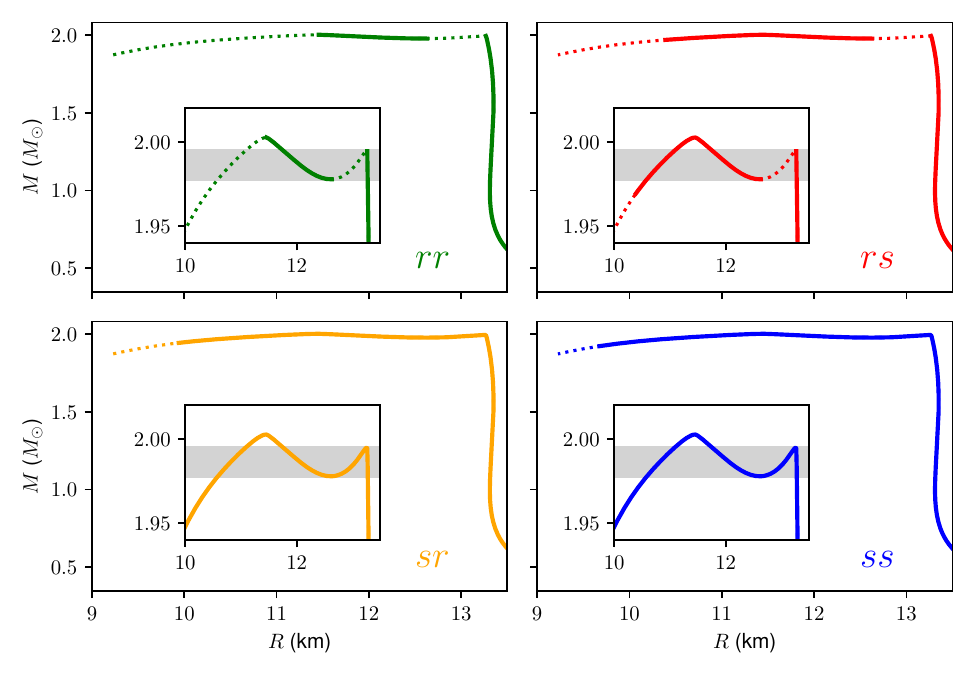}
\caption{$M$--$R$ diagrams for EoS parametrization A for the four different phase transition speed configurations. Stable/unstable regions are shown with solid/dotted lines, and the classical twin-supporting mass range is shaded.}
\label{Fig:MvsRQuadruplet}
\end{figure}

\begin{figure}
\includegraphics[width=\columnwidth]{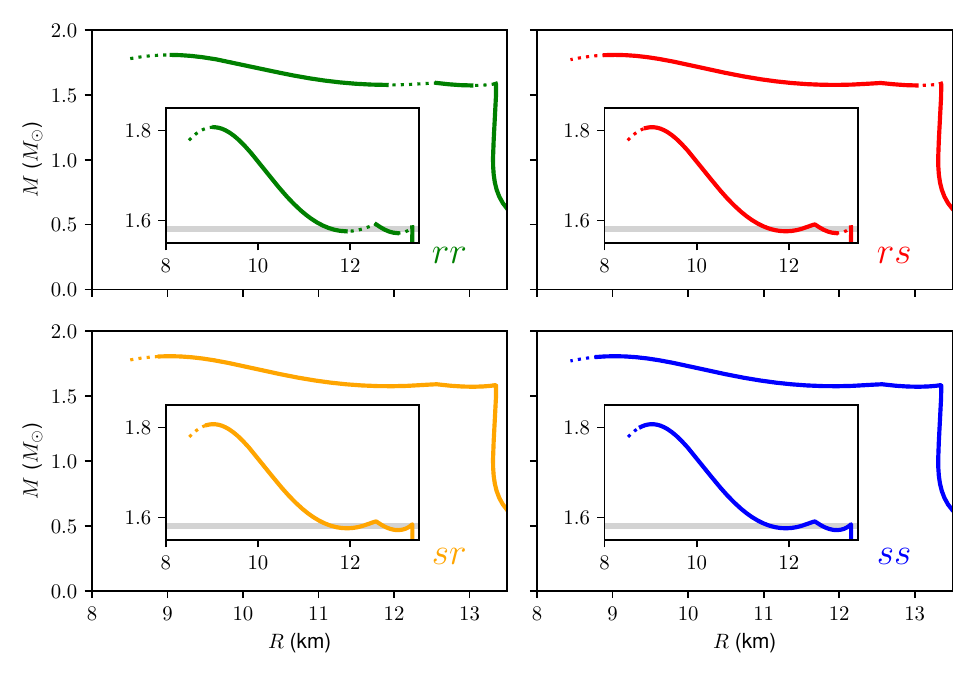}
\caption{Same as Figure~\ref{Fig:MvsRQuadruplet} but for EoS parametrization B.}
\label{Fig:MvsRQuintuplet}
\end{figure}

\begin{figure}
\includegraphics[width=\columnwidth]{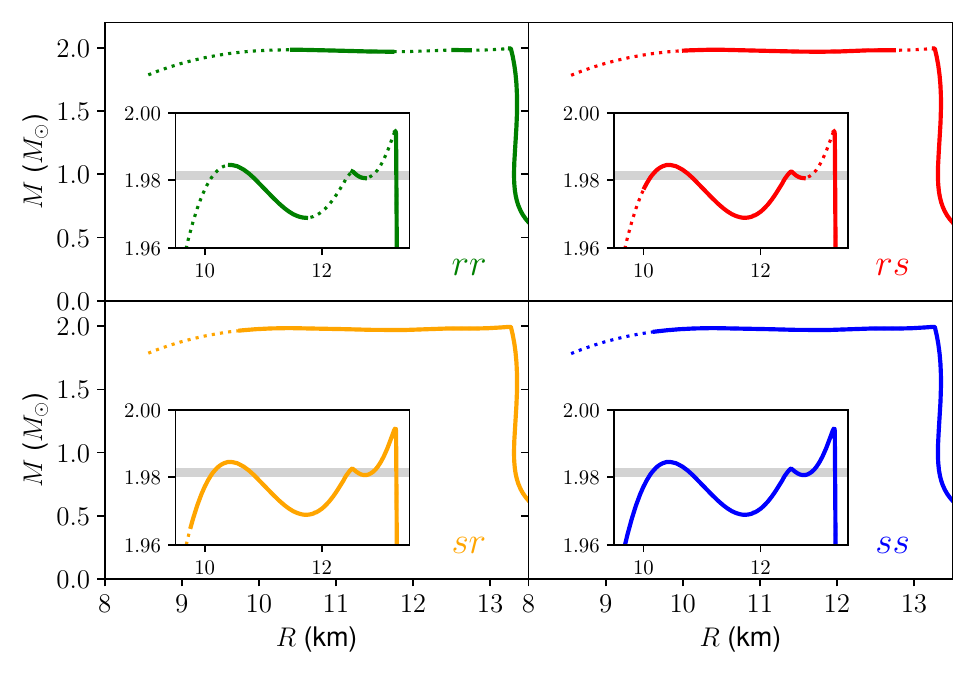}
\caption{Same as Figure~\ref{Fig:MvsRQuadruplet} but for EoS parametrization C. The classical triplet-supporting mass range is shaded.}
\label{Fig:MvsRSextuplet}
\end{figure}

We can use the information about stability provided by computing the fundamental radial modes to determine the ranges of radii for which stable stars exist given a particular configuration of phase transition speeds. The $M$ vs. $R$ diagrams for the four different phase transition speed configurations are given in Figures~\ref{Fig:MvsRQuadruplet}--\ref{Fig:MvsRSextuplet} for EoS parametrization A--B. This shows how different phase transition speed configurations at multiple first-order transitions could be distinguished from each other by filling in the $M$--$R$ diagram with observations, as each configuration has (sometimes subtle) differences in terms of the radius range which supports stable stars.

\begin{figure}
\includegraphics[width=\columnwidth]{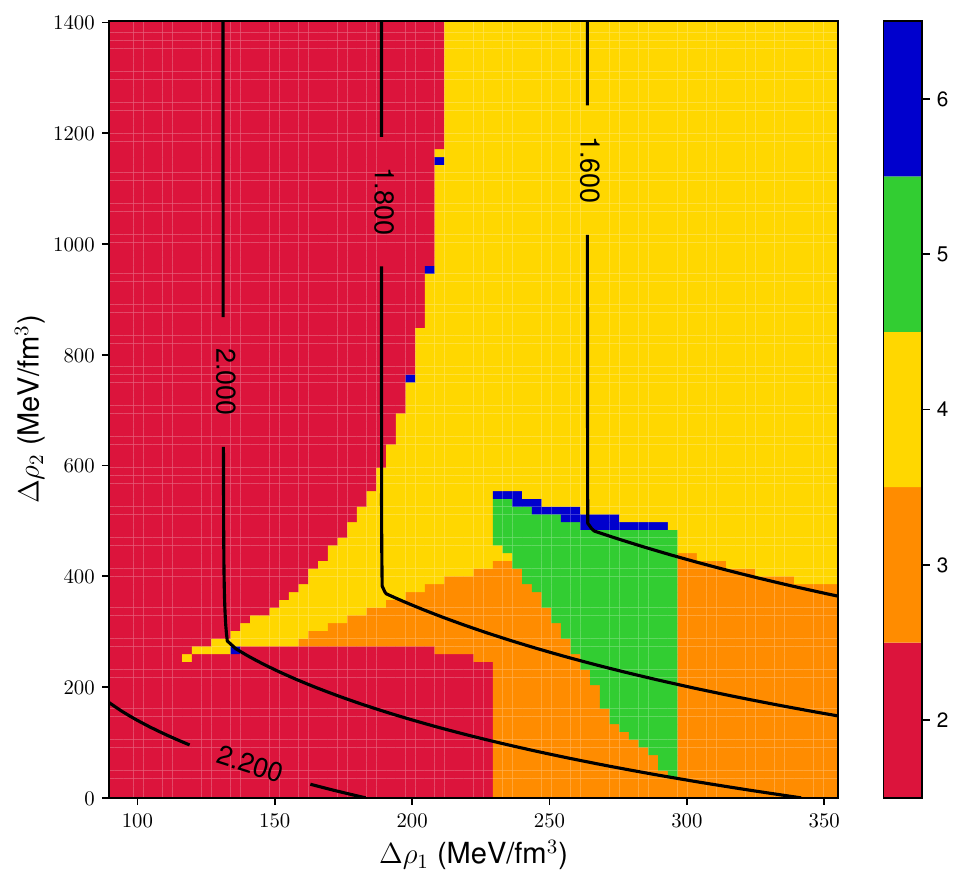}
\caption{Highest-order slow stable multiplet supported by a particular EoS parametrization, using parametrization B from Table~\ref{tab:EoSParameters} but with varying $\Delta\rho_1$ and $\Delta\rho_2$. Regions supporting slow stable twins are denoted by a 2, triplets by a 3, etc. The overlaid contours show the maximum mass supported by a particular EoS parametrization in units of $M_{\odot}$. The maximum value of $\Delta\rho_1$ is set by $\Delta\rho_1=\rho_2-\rho_1$. The region in parameter space for $0<\Delta\rho_1<90$ MeV/fm$^3$ (not shown) supports only slow stable twins. Note the small, isolated sextuplet-supporting regions along the boundary between the quadruplet and twin-supporting regions.}
\label{Fig:MultipletParameterSpace1}
\end{figure}

\begin{figure}
\includegraphics[width=\columnwidth]{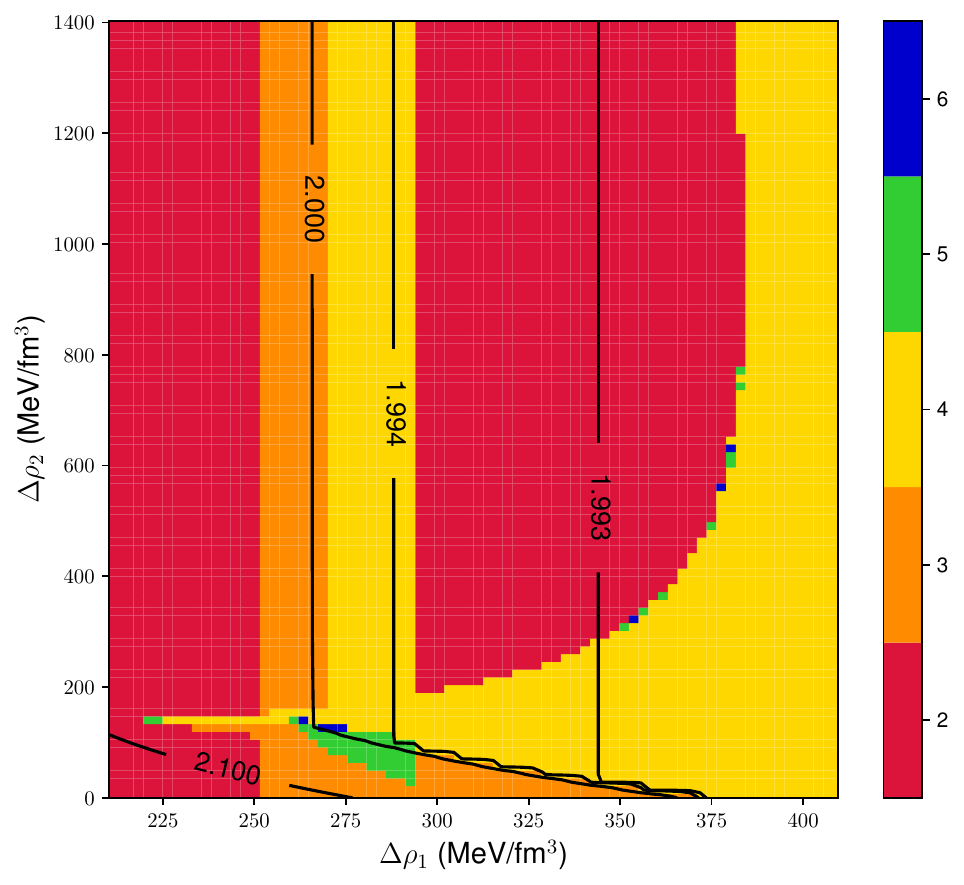}
\caption{Same as Fig.~\ref{Fig:MultipletParameterSpace1} but using parametrization C from Table~\ref{tab:EoSParameters} but with varying $\Delta\rho_1$ and $\Delta\rho_2$. The region in parameter space for $0<\Delta\rho_1<210$ MeV/fm$^3$ (not shown) supports only slow stable twin stars. Note the small, isolated quintuplet or sextuplet-supporting regions along the boundary between the quadruplet and twin-supporting regions.}
\label{Fig:MultipletParameterSpace2}
\end{figure}

How rare are EoS parametrizations which can support slow stable sextuplets compared to those which can support slow stable quadruplets and quintuplets? Figures~\ref{Fig:MultipletParameterSpace1}--\ref{Fig:MultipletParameterSpace2} show the regions in $\Delta\rho_1$--$\Delta\rho_2$ parameter space which support different slow stable multiplets for the $ss$ configuration. The other EoS parameters are held constant with their values in Table~\ref{tab:EoSParameters} column B (Fig.~\ref{Fig:MultipletParameterSpace1}) and column C (Fig.~\ref{Fig:MultipletParameterSpace2}). The overlaid contour lines show the maximum mass stellar model for the specified EoS parametrization. Only the highest-order multiplet is shown, so if e.g., an EoS parametrization supports multiple sets of slow stable twins in different mass ranges, this parametrization would be labeled ``2''. For both parametrizations, slow stable twins are the most common multiplet, followed by quadruplets, then triplets.

In Fig.~\ref{Fig:MultipletParameterSpace1}, quintuplets exist mostly within a wedge-shaped region spanning roughly $230\lesssim\Delta\rho_1\lesssim295$ MeV/fm$^3$ and $50\lesssim\Delta\rho_2\lesssim550$ MeV/fm$^3$. For greater $\Delta\rho_2$ only quadruplets are supported, with a narrow boundary region in parameter space supporting sextuplets between the quadruplet and quintuplet regions. Like in Fig.~\ref{Fig:MultipletParameterSpace1}, in Fig.~\ref{Fig:MultipletParameterSpace2} quintuplets exist mostly within a wedge-shaped region but which spans a smaller region of parameter space given by roughly $260\lesssim\Delta\rho_1\lesssim290$ MeV/fm$^3$ and $30\lesssim\Delta\rho_2\lesssim150$ MeV/fm$^3$. Sextuplets are similarly possible in a very narrow region of parameter space along the boundary between the quintuplet and quadruplet-supporting regions, largely concentrated around $\Delta\rho_1=270$ MeV/fm$^3$ and $\Delta\rho_2=150$ MeV/fm$^3$. Quintuplets or sextuplets are also possible in a very narrow region along the boundary between the quadruplet and twin-supporting regions for both parametrizations-- these regions can be difficult to notice in Fig.~\ref{Fig:MultipletParameterSpace1} and~\ref{Fig:MultipletParameterSpace2}. The regions of parameter space allowing triplets and quadruplets are much larger in Fig.~\ref{Fig:MultipletParameterSpace1} than Fig.~\ref{Fig:MultipletParameterSpace2}. Another significant difference between these parametrizations is the maximum mass: for B, the maximum mass drops from above $2M_{\odot}$ to below $1.6M_{\odot}$ as $\Delta\rho_1$ is increased, while for C it is at or above (within $<1\%$) $2M_{\odot}$ for the entire $\Delta\rho_1$--$\Delta\rho_2$ parameter space.

\section{The reaction mode in stars with multiple first-order phase transitions}
\label{sec:ReactionMode}

The reaction mode is a distinct class of radial oscillation that only appears in the mode spectrum for a compact star with a rapid phase transition. This mode is distinguished by examining the radial mode spectrum in the limit of vanishing higher-density phase core;  the reaction mode does not correspond to any mode of a single-phase star, and so there should be a frequency jump when going from one phase to another for the same radial node number mode. The reaction mode is often the fundamental mode, but it can be an overtone. In effect this mode slots into the usual mode spectrum, pushing the other modes to a higher frequency than in the slow phase transition case.

\begin{figure}
\centering
\includegraphics[width=\linewidth]{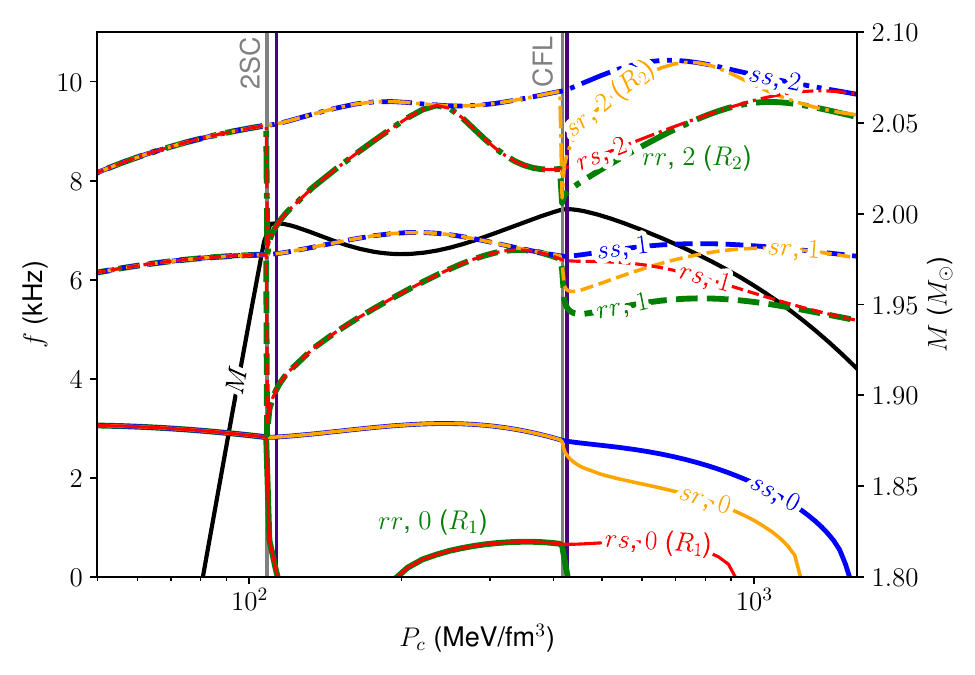}
\caption{Fundamental and first two overtones radial modes for EOS parametrization A, with the corresponding stellar mass shown on the secondary vertical axis. Modes are denoted by a pair of letters $r$ and $s$ indicating the rate of the phase transitions, and a number indicating the fundamental (0), the second harmonic (1), or third harmonic (2). The reaction modes for the two phase transitions are labeled with an additional $R_1$ or $R_2$ denoted the reaction mode for the nuclear--2SC and 2SC--CFL phase transition respectively.}
\label{Fig:ReactionMode1}
\end{figure}

\begin{figure}
\centering
  \subfloat[]{\includegraphics[width=\linewidth]{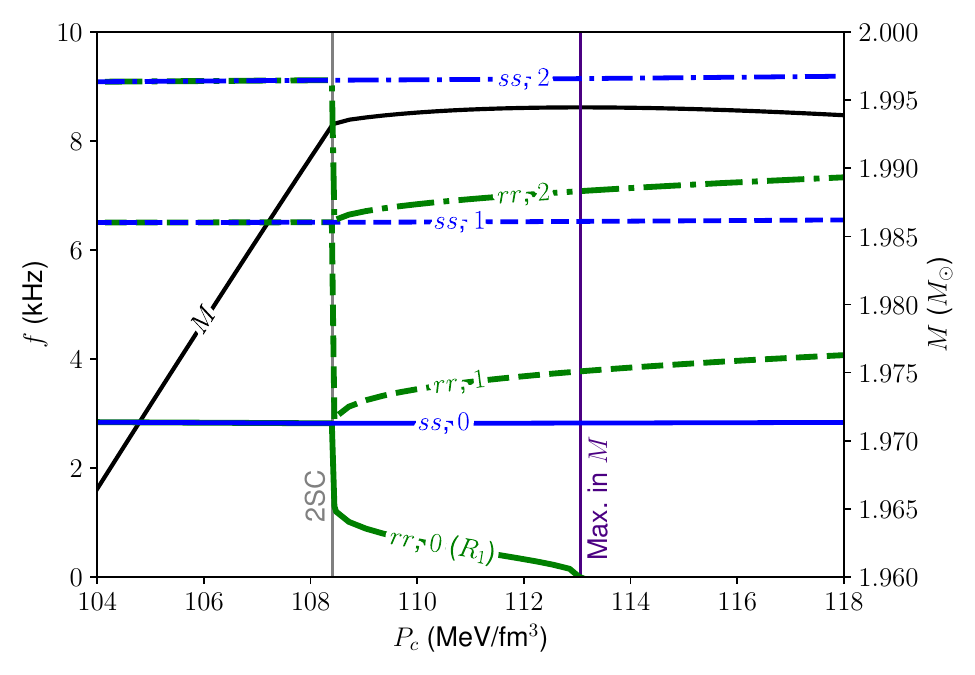}}
  \vfill
  \subfloat[]{\includegraphics[width=\linewidth]{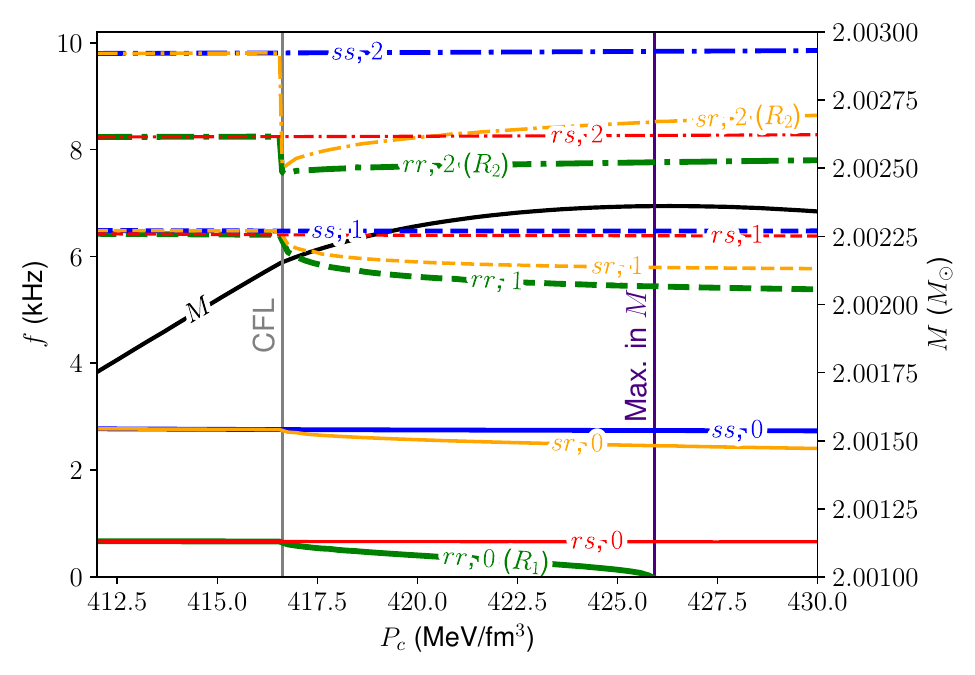}}
\caption{(a) Zoomed-in view of FIG.~\ref{Fig:ReactionMode1} near the nuclear--2SC phase transition and the first local maximum in $M(P_c)$. Note that the $rr$ and $rs$, and $ss$ and $sr$, configurations are identical between the two phase\
 transitions. (b) Zoomed-in view of FIG.~\ref{Fig:ReactionMode1} near the 2SC--CFL phase transition and the second local maximum in $M(P_c)$.}
\label{Fig:ReactionMode1Zoomed}
\end{figure}

To demonstrate the changes to the reaction mode in hybrid stars with multiple first-order phase transitions of different transition rates, we examine the mode spectrum of stars for different phase transition configurations and EoS parametrizations. First,  Fig.~\ref{Fig:ReactionMode1} shows the first three radial modes for EoS parametrization A and the four different phase transition rate configurations. Fig.~\ref{Fig:ReactionMode1Zoomed} focuses on the phase transitions and local maxima in $M(P_c)$ to show the behavior of the $rr$ modes in particular in this region. For this EoS parametrization, panel (a) shows the fundamental mode is the reaction mode for a rapid nuclear--2SC phase transition ($rr$ and $rs$). 

For the same EoS parametrization, examining the 2SC--CFL phase transition in Fig.~\ref{Fig:ReactionMode1Zoomed} panel (b) shows the novel behavior of the reaction mode for a two phase transition star since both $rr$ and $sr$ configurations have a reaction mode here. We label this second reaction mode $R_2$; it is the first (and only) reaction mode for the $sr$ model but the second such mode for the $rr$ configuration. Note first that the fundamental mode for the $rr$ configuration, which was the reaction mode $R_1$ for the nuclear--2SC phase transition, is \textit{not} the reaction mode for the 2SC--CFL phase transition since there is no frequency jump for the fundamental mode across the phase transition. It is now the third harmonic that is the reaction mode, for both $rr$ and $sr$ configurations. 

\begin{figure}
\includegraphics[width=\columnwidth]{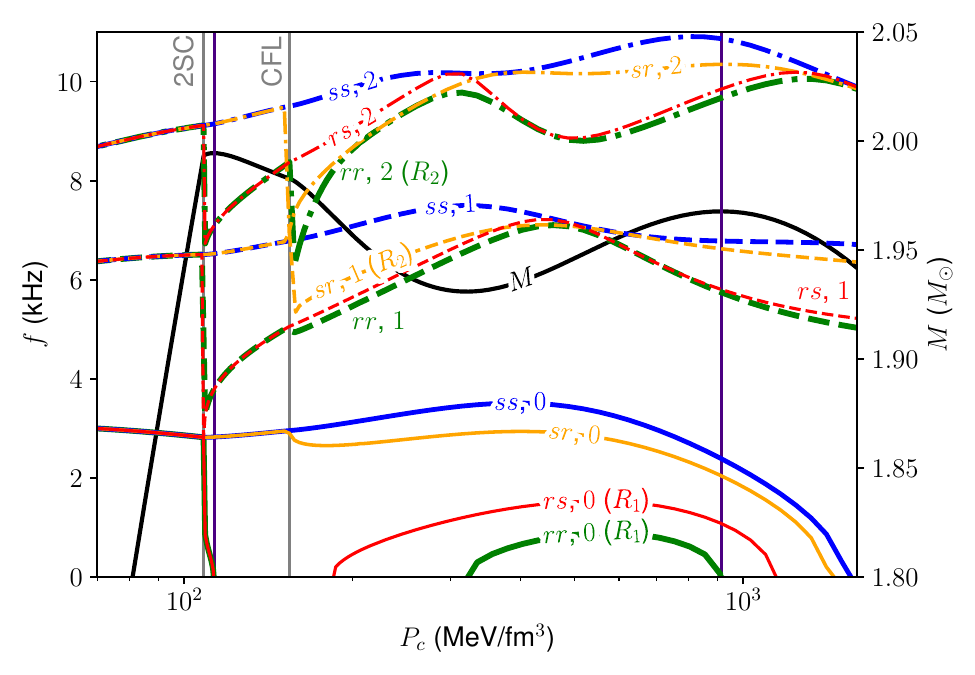}
\caption{Fundamental and first two overtones radial modes for EoS parametrization D, with the corresponding stellar mass shown on the secondary vertical axis. Other labelling follows Fig.~\ref{Fig:ReactionMode1} (a).}
\label{Fig:ReactionMode2}
\end{figure}

Using a different EoS parameterization, we can show another interesting aspect of the $R_2$ modes. The fundamental radial mode and first two overtones for the EoS parametrization D in Table~\ref{tab:EoSParameters} are shown in Fig.~\ref{Fig:ReactionMode2}. In this case, the frequency of the $R_2$ mode in the limit of zero CFL core can differ between the $rr$ and $sr$ configurations, and the corresponding overtone is different: for the $sr$ case it is the second harmonic, and for the $rr$ case the third harmonic. Note that the third harmonic in the limit of zero CFL core converges to the second harmonic for the nuclear plus 2SC $s$ configurations; the second harmonic is pushed to higher frequencies, corresponding to a higher radial node number and hence higher overtone, by the reaction mode.

\section{Choices of junction physics}
\label{sec:Junctions}

The junction conditions employed in previous calculations of radial oscillations of two or more phase compact stars have been the rapid and slow conditions given by Eq.~(\ref{eq:SlowConversionJunctions}) and~(\ref{eq:RapidConversionJunctions}) respectively. However, these are not the only possible conditions, and in fact~\cite{Karlovini2004} found the most general junction conditions at first-order phase transitions for the radial perturbations of general relativistic fluids. These conditions are
\begin{subequations}
\begin{align}
\left[\xi-\mathcal{F}\right]^+_-={}&0,
\label{eq:GenJunction1}
\\
\left[(\rho+P)\mathcal{F}\right]^+_-={}&0,
\label{eq:GenJunction2}
\\
\left[\Delta P\right]^+_-={}&0.
\label{eq:GenJunction3}
\end{align}
\end{subequations}
where $$\mathcal{F}=\frac{\Delta F}{r}\left(\frac{\text{d}F}{\text{d}r}\right)^{-1},$$ for a function $F=F(r)$ which defines the phase boundary. The rapid conversions correspond to $F=P$, and the slow conversions to taking $\Delta F=0$. While these two limits are the most obvious, and simplest, choices of junction conditions and have simple physical interpretations, there are other possible choices. So an interesting question is: could alternative choices of $F$ change stability, perhaps restoring the BTM criterion by destabilizing the slow stable stellar models? We examine alternative junction conditions in the following subsections.

\subsection{Phase transition at fixed radius}

\begin{figure}
\includegraphics[width=\columnwidth]{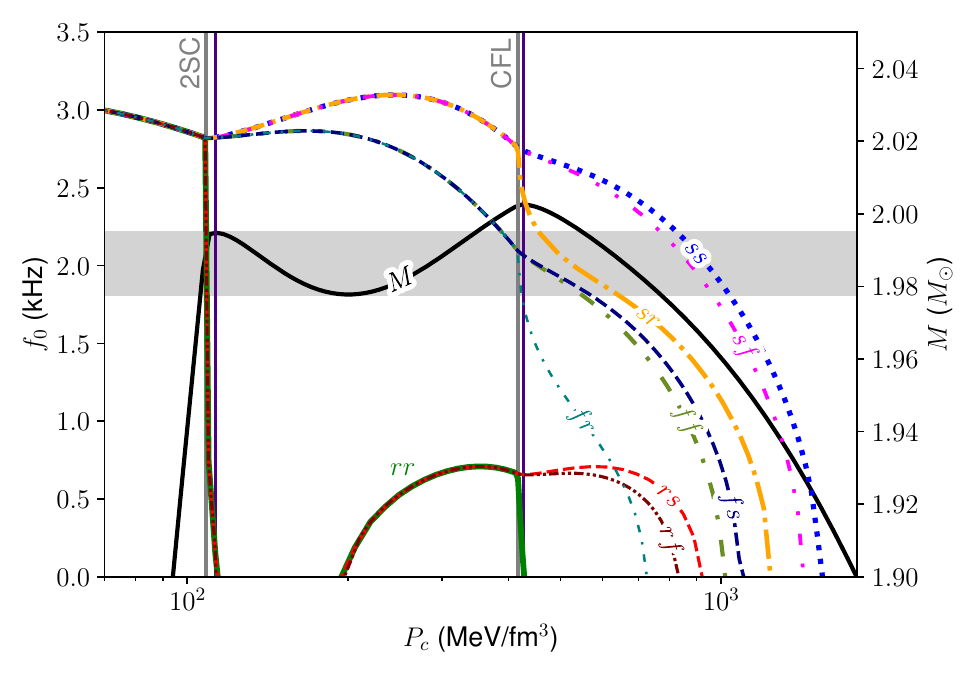}
\caption{Fundamental radial modes for EoS parametrization A, including all nine permutations of the phase transition junction conditions including the fixed radius junction conditions Eq.~(\ref{eq:FixedRadiusJunctionConditions}) (denoted $f$ in the line labeling).}
\label{Fig:FixedRadiusFundamentalModes}
\end{figure}

As an alternative junction condition to the usual slow and rapid cases, we examine the stability implications of simply taking $F=r$ and thus $\mathcal{F}=\xi$, which corresponds to the phase transition occurring at a fixed radius. Eq.~(\ref{eq:GenJunction1}) reduces to $0=0$ in this case, and the two nontrivial junction conditions are now
\begin{equation}
\left[(\rho+P)\xi\right]^+_-=0, \qquad \left[\Delta P\right]^+_-=0.
\label{eq:FixedRadiusJunctionConditions}
\end{equation}
The first condition corresponds to the continuity of the stress-energy tensor $T_{\mu}^{\ \nu}$ projected perpendicular to the phase transition boundary. Denoting a phase transition with this junction condition with the letter $f$, there are now nine possible configurations of the two phase transitions. The fundamental radial modes for each of these combinations are plotted in Fig.~\ref{Fig:FixedRadiusFundamentalModes} using EoS parametrization A. The fixed radius junction condition stabilizes the fundamental mode in the BTM unstable regions like the slow conversion rate junction condition, but the fundamental modes when this junction condition is 
used becomes unstable at a lower $P_c$ than the corresponding mode calculated using the slow conversion rate condition.

\subsection{Intermediate speed phase transition}

For a barotropic EoS $P=P(\rho)$, $F$ can only be a function of $P$ or $r$, and any attempt to write $F$ as a more complicated function of $P$ will give $\mathcal{F}$ that is equal to the rapid conversion case. As an attempt to determine the stability implications of an intermediate conversion speed between the slow and rapid conversion speed cases, we thus choose junction conditions that do not follow from Eq.~(\ref{eq:GenJunction1}) for any particular $F$, but interpolate between the slow and rapid cases. Thus the following junction conditions can be selected:
\begin{equation}
\left[\Delta P\right]^+_-=0, \qquad \left[\xi-\alpha\frac{\Delta P}{r}\left(\frac{\textrm{d}P}{\textrm{d}r}\right)^{-1}\right]^+_-=0,
\label{eq:IntermediateSpeedJunctionConditions}
\end{equation}
where the value of $0\leq\alpha\leq 1$ is varied. The slow and rapid conversion rate junction conditions are recovered  by choosing $\alpha=0$ and $\alpha=1$, respectively. Note that Eq.~(\ref{eq:IntermediateSpeedJunctionConditions}) are not proper junction conditions according to Eq.~\eqref{eq:GenJunction1}--\eqref{eq:GenJunction3}, and we are only using them as a way to smoothly interpolate between two cases which do individually satisfy Eq.~\eqref{eq:GenJunction1}--\eqref{eq:GenJunction3}.

\begin{figure}
\includegraphics[width=\columnwidth]{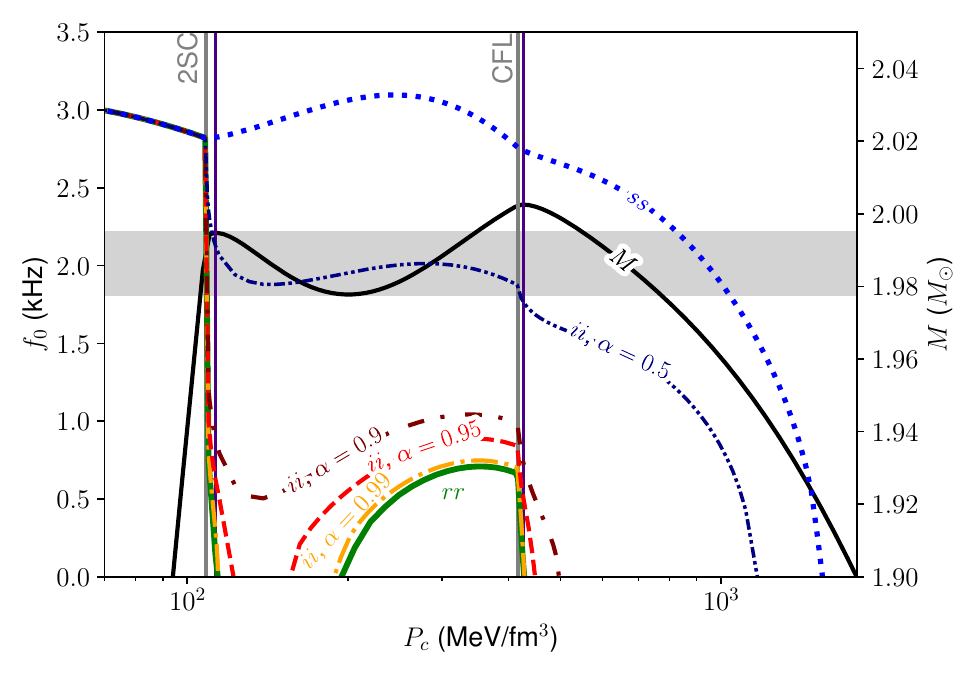}
\caption{Fundamental radial modes for EoS parametrization A, including the $rr$ and $ss$ phase transition speed configurations and the $ii$ (two intermediate conversion speed transitions) configuration for different values of the parameter $\alpha$ defined in Eq.~(\ref{eq:IntermediateSpeedJunctionConditions}).}
\label{Fig:IntermediateSpeedFundamentalModes}
\end{figure}

Denoting the intermediate conversion speed junction condition by an $i$, we plot the fundamental radial mode for EoS parametrization A with two intermediate conversion speed phase transitions in Fig.~\ref{Fig:IntermediateSpeedFundamentalModes}. Varying values of $\alpha$ are used, showing that for $\alpha>0$, there is at least some infinitesimally small set of central pressures for which the star is stabilized in the $ii$ configuration that would be unstable in the $rr$ configuration. Even a moderate value of $\alpha=0.9$ stabilizes the stellar models with central pressures spanning the entirety of the 2SC phase. The main distinguishing characteristic of smaller values of $\alpha$ in terms of stability is thus the central pressure beyond which all stars are unstable, which approaches the $ss$ value for $\alpha\rightarrow 0$.

Both the fixed radius and intermediate conversion speed junction conditions share similarities with the slow conversion speed conditions in terms of their stability implications; they can both stabilize stellar models that are unstable according to the BTM criterion and in the $rr$ configuration. This suggests that even if the physics of these junction conditions is not exact and more complicated microphysical input is required to accurately represent them, it should be expected that at least certain stellar models which violate the BTM criterion will be stable if they contain a strong first-order phase transition.

\section{Conclusion}
\label{sec:Conclusion}

Studying compact star stability and the implications that phase transitions have on this could allow us to determine properties of phase transition in dense matter from observations of compact stars. In particular, the existence of twin stars or higher-order stellar multiplets would strongly suggest the presence of a first-order phase transition in the star's core, possibly between hadronic and deconfined quark matter. It is also theoretically possible that multiple phase transitions could occur within the density range of a very massive compact star's core.  Motivated by this we studied the stability of hybrid stars with multiple first-order phase transitions from nuclear to quark matter and between quark matter phases using a simple parametrized EoS. The stability of the compact stellar models was determined by computing the fundamental radial oscillation modes of the stars while imposing different junction conditions on the oscillation modes at the phase transitions. These junction conditions, and the rate of the conversion between phases at the phase transitions, can have significant effects on the stability as has been demonstrated in previous calculations.

We demonstrated the existence of higher-order stellar multiplets up to sextuplets with two slow first-order phase transitions. The members of a multiplet higher than a triplet are slow stable stars that would be classified as unstable according to the BTM criterion but are stable according to their fundamental radial mode having a real frequency. However, we found that stable sextuplets with slow stable stars only exist for a narrow range of EoS parameter space, and require that the maximum of the $M$--$P_c$ curve occurs at pressures close to that of the 2SC--CFL phase transition. Slow stable quadruplets and quintuplets are far more common in the EoS parametrizations we examined, and likely in general.

We also studied the reaction modes, the radial modes only existing in stars with rapid phase transitions, for a two phase transition compact star for the first time. We showed that there exists one reaction mode per rapid phase transition and that the overtone of the reaction mode is not necessarily the same for each phase transition (e.g., the fundamental mode is the reaction mode for the first phase transition and the second harmonic is the reaction mode for the second phase transition). We also discussed the stability implications of alternative junction conditions for the radial oscillation modes at the phase transitions in compact stars. We examined two simple alternative sets of junction conditions corresponding to the phase transition happening at a fixed radius and interpolation between the slow and rapid conversion speed cases. For both of these alternatives, the resulting radial modes and stability behavior are closer to the slow conversion speed junction conditions than the rapid conversion speed conditions.

The conclusions of this paper were reached using a simple sequential constant speed of sound parametrization of the EoS of quark matter that can describe a three-phase compact star core, but an improved study could employ a wider variety of EoS. As constructed, the EoS
for example does not recover the conformal limit of the speed of sound 
$c/\sqrt{3}$ from below~\cite{Bedaque2015}. To satisfy this, one needs a non-monotonic dependence of sound speed with density which has been proposed as necessary to support $2M_{\odot}$ {\it and} satisfy the conformal limit~\cite{Tews2018a}. EoSs which include more microphysics or, for example, are non-barotropic (e.g., including dependence on species fractions like the proton fraction in nuclear matter or strangeness in quark matter), could be better suited for modeling complex phase transitions, especially at non-zero temperatures which are relevant for binary neutron star mergers. Such approaches may allow for more complicated junction conditions.  As we have shown, employing the correct junction conditions is key to determining the existence of $s$ hybrid stars and their corresponding higher-order stable multiplets.

\section*{Acknowledgements}

We thank the referees for helpful comments. P.~B.~R. was supported by the Institute for Nuclear Theory's U.S. Department of Energy grant No. DE-FG02-00ER41132.  A.~S. acknowledges support by Deutsche Forschungsgemeinschaft Grant No. SE 1836/5-2 and the Polish NCN Grant No. 2020/37/B/ST9/01937.

\bibliographystyle{apsrev4-2}
\bibliography{library,librarySpecial,textbooks,Hybridstars_refs}

\end{document}